\documentstyle[aps,epsf]{revtex}  % DO NOT DELETE THIS LINE 

\begin{document}        %  DO NOT DELETE OR CHANGE THIS LINE

\baselineskip 14pt
\title{Measurement of the Neutrino-induced Semi-contained Events in
  MACRO}
\author{R. Nolty (For the MACRO Collaboration)}
\address{Lauritsen Laboratory of High Energy Physics\linebreak
  California Institute of Technology}
% \author{}   % Use this and the next line only if there is a second
% \address{Another University, etc.}  % address. (Remove the left % marks)
%
\maketitle              % Creates the title area, Do Not Remove

\begin{abstract}        % Do Not Delete this line
  A preliminary analysis is presented of low-energy ($\overline{E_\nu}
  \sim 4 GeV$) neutrino interactions observed by the MACRO detector.
  These include neutrino interactions in the detector, as well as
  upgoing neutrino-induced muons that stop in the detector.  At these
  energies, essentially all observed interactions can be attributed to
  atmospheric neutrinos.  A large deficit is observed compared to a
  no-oscillations Monte Carlo prediction, although the systematic
  errors are large.  However, the observation agrees well with
  neutrino oscillations with parameters suggested by the MACRO
  upward throughgoing muon analysis as well as other experiments
  (maximal mixing with $\Delta m^2$ of a few times $10^{-3}$).  \ 
\end{abstract}          % Do Not Delete this line

\section{Introduction}               % Introduction goes below.

Recent measurements of atmospheric neutrino flux \cite{SuperK},
\cite{Soudan} as well as some older measurements \cite{IMB},
\cite{Kamiokande} give absolute values and ratios of various
quantities that are inconsistent with expectations for massless
(non-oscillating) neutrinos, although some older measurements
\cite{Baksan}, \cite{Frejus}, \cite{NUSEX}, \cite{IMBStop} reported no
evidence for oscillations.  The ongoing MACRO measurement \cite{Nat}
of neutrino-induced upgoing muons that traverse the entire detector
(so-called ``throughgoing muons'') also suggests oscillations with
parameters ($\Delta m^2$ a few times $10^{-3}$ and $sin^2 2\theta =
1$) roughly consistent with \cite{SuperK} and \cite{Soudan}.  The
MACRO analysis has been extended to event topologies that probe lower
neutrino energies, and the preliminary results are presented here.

\begin{figure}[bth]      % in second brace, h=here, t=top, b=bottom      
%\centerline{\epsfxsize 1.1 truein \epsfbox{sctop.eps}}   
  \centerline{\epsfxsize 3.7 in \epsfbox{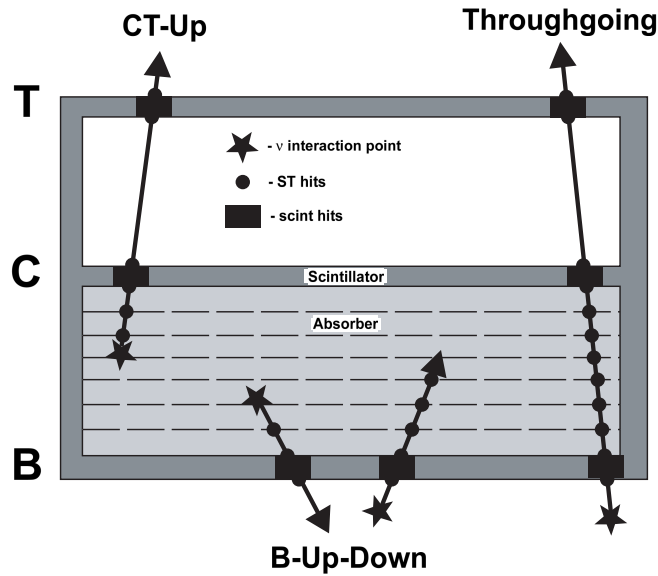}}   
  \vskip -.2 cm
  \caption[]{
    \label{fig:sctop}
    \small Event topologies of neutrino-induced events in MACRO.}
\end{figure}

The MACRO detector \cite{MACROTech} located at a depth of 3700 mwe at
the Gran Sasso laboratory in Italy is a large (77 m $\times$ 12 m
$\times$ 9.3 m) detector of penetrating radiation.  Although it was
designed primarily to search for exotic slow-moving supermassive
particles such as GUT monopoles, it is also capable of measuring
neutrino-induced muon fluxes.  The bottom half of the detector is
filled with crushed rock absorber and planes of limited streamer tubes
(with wire and strip views) with a pitch of 3 cm.  Each outer face of
the detector and a central layer are filled with boxes of liquid
scintillator that provide sub-nanosecond timing resolution.  The outer
faces also contain additional streamer tube layers.  The interior of
the upper portion of the detector is hollow.  (See
Figure~\ref{fig:sctop}.)

\begin{figure}[ht]      % in second brace, h=here, t=top, b=bottom      
  \centerline{\epsfxsize 3 truein \epsfbox{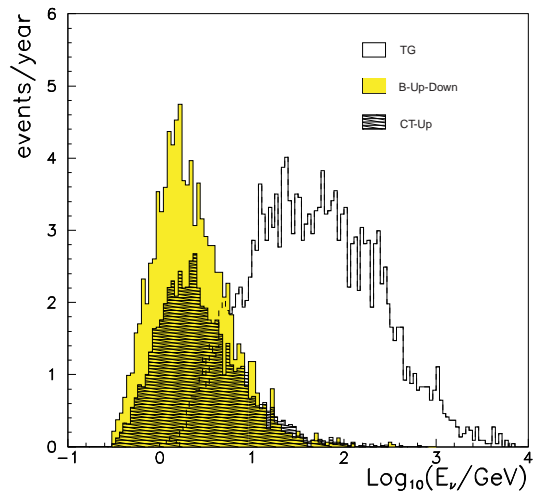}}   
  \vskip -.2 cm
  \caption[]{
    \label{fig:scnrg}
    \small Neutrino energy distributions giving rise to the different
    event topologies in MACRO.}
\end{figure}

Almost all neutrino interactions in the detector take place in the
lower half, where most of the mass is.  Because of its large
granularity, MACRO has very poor efficiency to detect neutral current
events or neutrino-induced electrons.  It is primarily sensitive to
muons (from $\nu_{\mu}$ charged current interactions) which travel a
few meters or more.  There are three event topologies of
neutrino-induced muons analyzed in MACRO (again, refer to
Figure~\ref{fig:sctop}).  Throughgoing muons, labeled TG, are from
neutrino interactions below the detector which send a muon through the
entire detector.  The TG analysis is described elsewhere~\cite{Nat}.
Upward contained-vertex events, labeled CT-Up, in which the muon
strikes the Center (``C'') layer of scintillator as well as a higher
layer (``T'' for Top) constitute the first topology in the present
analysis.  Of course, there are also neutrino-induced downgoing
muons, both throughgoing and stopping in the detector.  However, they
are indistinguishable from primary atmospheric muons from cosmic ray
showers, which even at MACRO's depth outnumber neutrino-induced muons
by 100,000 to 1.  Therefore for these topologies only upward muons are
considered, as determined by the time of flight between two or three
scintillator layers.  The final topology, labeled B-Up-Down, consists
of two classes of events: downward muons from contained-vertex events
and upward muons from external events which stop in the detector.
Because only one layer of scintillator is hit (``B'' for Bottom) and
because the streamer tubes do not provide accurate timing information,
it cannot be determined if the particles are moving up or down, so the
B-Up-Down analysis sums over both classes of events (upward and
downward).  Figure~\ref{fig:scnrg} shows the distribution of parent
neutrino energies contributing to the different topologies, from a
no-oscillations Monte Carlo calculation (described below).  The
current analyses utilize neutrinos more than an order of magnitude
less energetic than the TG analysis.

\section{The CT-Up Analysis}

The first analysis, CT-Up, looks for upward muons from neutrino
interactions in the lower part of the detector.  The muon must strike
a scintillator tank in the center layer, as well as a higher layer
(either the top layer or the upper portion of the side walls).  The
analysis requires these two scintillator hits, as well as enough
colinear streamer tube hits in two different views to reconstruct a
track in space.  Typically this is 4 hits per view, but varies
depending on which planes the track traverses.  To ensure that the
vertex is truly internal, a projection of the track below the lowest
streamer tube hit is required to pass through several streamer tube or
scintillator planes which did not fire.  Thus, muons that enter from
below through cracks in the detector are rejected.

\begin{figure}[ht]      % in second brace, h=here, t=top, b=bottom      
  \centerline{\epsfxsize 3 truein \epsfbox{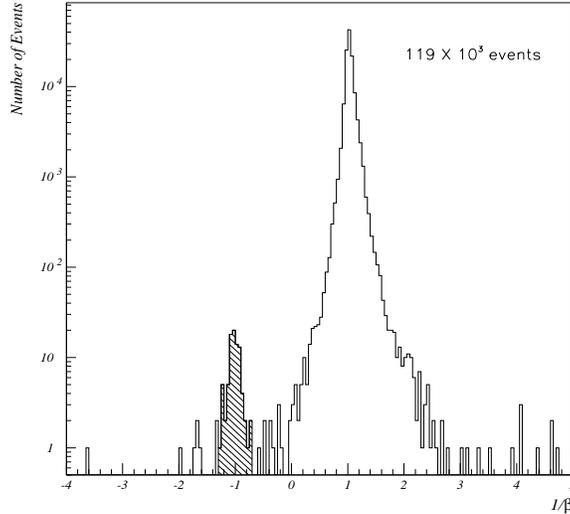}}   
  \vskip -.2 cm
  \caption[]{
    \label{fig:beta}
    \small $1/\beta$ distribution of events in the CT-Up analysis.
    Downgoing events are clustered about $\beta = 1$ while upgoing
    events are near $\beta = -1$.}
\end{figure}

The vast majority of events passing these geometry cuts are downgoing
atmospheric muons.  The final cut relies on the time of flight
measured between the two scintillator boxes to select only upgoing
events.  Defining $\beta$ by $v = \beta c$, with the convention
that upgoing particles have negative $\beta$, Figure~\ref{fig:beta}
shows the observed $1/\beta$ distribution for all events passing the
geometry cuts.  A substantial peak of upgoing events is well-separated
from the large background of downgoing events, although the
events between the peaks show a residual background due to mistimed
events.

\section{The B-Up-Down Analysis}

The B-Up-Down analysis searches for two types of events, both of which
have the same topological signature in MACRO: downward contained
vertex events and externally-produced upward muons that stop in the
detector.  In both cases the signature is a hit in the bottom
scintillator layer and a few associated colinear streamer hits.  In
the absence of oscillations, we expect about equal numbers from the
two classes because for every downgoing neutrino that makes a track
starting in the detector and ending below, there is a corresponding
upgoing neutrino that could make a track of the same length in the
opposite direction.

The analysis requires a B-layer scintillator hit associated with
streamer tube tracks in two views (wire and strip).  At least 3
colinear hits are required to define a track, which implies at least
$100 g/cm^2$ of scintillator and crushed rock must be traversed.  To
reduce the background attributable to the huge number of downgoing
primary atmospheric muons, all hits are required to be more than $1 m$
from the side walls, and a projection of the track above the highest
streamer tube hit must pass through several streamer planes and/or
scintillator tanks which did not fire.

Experience has shown that many events passing these simple cuts do not
appear to be clean neutrino-induced events.  Therefore, a final hand
scanning procedure is implemented to reject events where the
reconstructed track appears wrong or hits outside the fiducial volume
appear to be correlated with the track.  Fully half the candidates are
rejected by the hand scan.  However, it should be noted that two
different people performed the scan and made the same judgment on more
than 95\% of events.  Simulated events passed the hand scan with
greater than 95\% efficiency.  A small systematic uncertainty is
included in the final results due to these effects.

In addition to neutrino-induced events, upgoing particles (typically
soft pions) are induced by downward atmospheric muons, through a
photonuclear process such as

\vskip 1 em
$\mu N \rightarrow \mu \pi X$

\begin{figure}[ht]      % in second brace, h=here, t=top, b=bottom      
  \centerline{\epsfbox{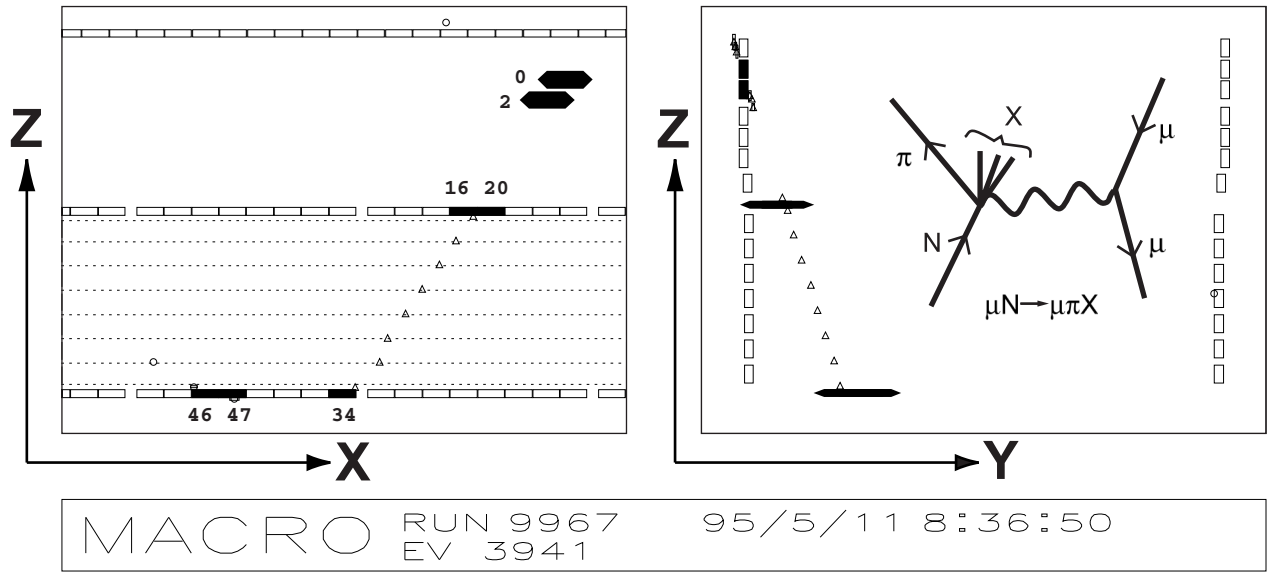}}
  \vskip -.2 cm
  \caption[]{
    \label{fig:bounce}
    \small A soft upward pion produced by a downward muon.  The
    numbers in the X-Z view give the relative time of scintillator
    hits in nanoseconds.  The downgoing muon traversed the detector
    for the first 34 ns, interacted below the detector, and the pion
    reentered the detector at 46 ns.}
\end{figure}

\vskip 1 em
\noindent
Figure~\ref{fig:bounce} shows an event probably produced by this
mechanism.  A downgoing muon traversed the detector, and a few
nanoseconds later an upgoing particle hit the detector nearby.  In
this case, because we saw the muon we would reject this event in the
neutrino analysis.  However, if the muon missed the detector and we
saw only the soft pion, it would be indistinguishable from the
neutrino-induced muons we are searching for.  We have made a
study~\cite{bounce} of 243 events similar to that shown in
Figure~\ref{fig:bounce} to characterize the spectrum of particles
produced by downgoing muons. We estimate such events give an
irreducible background which amounts to about 4\% of the number of
events we observe.  MACRO's non-compact geometry is well-suited for
measuring this background, and the results in~\cite{bounce} can be
adapted for use in other underground experiments.

\section{The Monte Carlo Calculation}

An approximate event rate could be determined by a semi-analytic
calculation; however, it is most precise to use a Monte Carlo
calculation that takes into account the detector geometry and all
relevant energy loss mechanisms.  The basic ingredients of the Monte
Carlo calculation are a calculation of the atmospheric neutrino flux,
a model of the neutrino interaction cross section, and detector
simulation.  We use the Bartol flux calculation~\cite{Bartol}
including geomagnetic effects.  The cross section model is due to
Lipari, et al~\cite{Lipari} which computes the total cross section as
the sum of three component processes -- quasielastic, resonant and
deep inelastic.  For the deep inelastic cross section, the current
work uses parton distribution function set S1 of Morfin and
Tung~\cite{Morfin}, but future work will utilize a more modern
distribution function.  The detector simulation is the standard
Geant-based MACRO simulation program, GMACRO, which has been tuned to
match the copious downgoing muon data.  Events output by GMACRO are in
the same format as real data, and are analyzed by the same software
chain as real data.

The calculation assumes that atmospheric neutrinos are the only
relevant source of neutrinos.  In this regard, we may point out that,
in the absence of oscillations, the atmospheric neutrino flux alone
overpredicts the number of observed events.  Also, MACRO searches for
two exotic sources of neutrino flux (WIMP annihilation in the center
of the earth or the sun~\cite{Doug}, and astrophysical point
sources~\cite{nuAstro}) have been negative.

In the case of the B-Up-Down analysis, after cuts Monte Carlo events
are randomly merged with real events before the hand scan, so that the
people performing the scan do not know if they are looking at a real
or a Monte Carlo event.

\section{Results}

\label{sec:results}

\begin{figure}[ht]      % in second brace, h=here, t=top, b=bottom      
  \centerline{\epsfxsize 6 truein \epsfbox{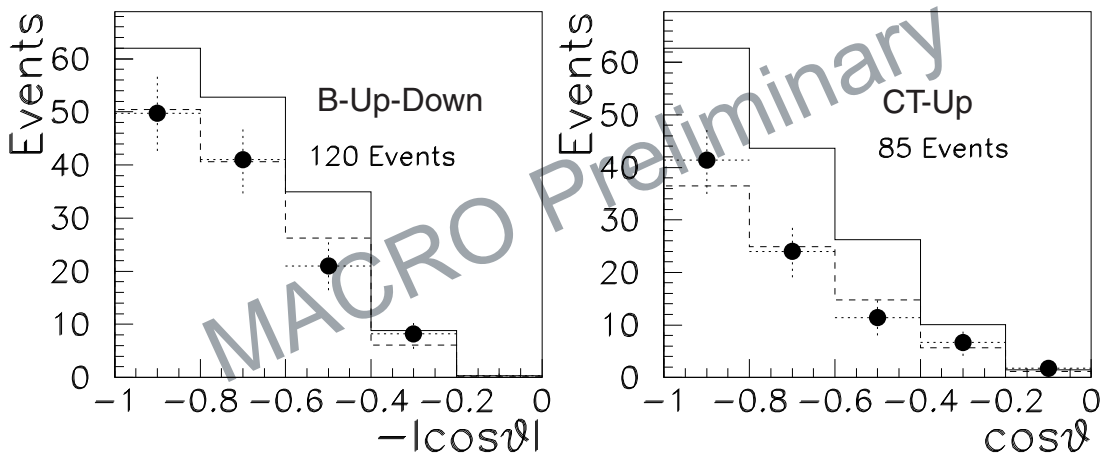}}   
  \vskip -.2 cm
  \caption[]{
    \label{fig:results}
    \small Zenith distributions of events selected in the two
    low-energy analyses.  The solid line gives the prediction of a
    no-oscillations Monte Carlo.  The dashed line gives the prediction
    for $\Delta m^2 = 2.5 \times 10^{-3}$ and $sin^2 2\theta = 1$.}
\end{figure}

The B-Up-Down analysis had an effective livetime of 2.81 years,
occurring between July, 1994 and November, 1997.  125 events passed
the scan, of which 5 are estimated to be background due to soft pion
production.  The zenith angle distribution of the
background-subtracted sample is shown in Figure~\ref{fig:results}a.
In this case, we do not know if any individual event is upgoing or
downgoing, so the horizontal axis is $-|cos\theta|$, with vertical to
the left and horizontal to the right.

The CT-Up analysis utilized a total effective livetime of 3.16 years,
accumulated between April, 1994 and November, 1997.  88 events are
identified in the upgoing peak of Figure~\ref{fig:beta}, of which an
estimated 3 are due to the mistimed background.  The zenith angle
distribution of the background-subtracted sample is shown in
Figure~\ref{fig:results}b.  All of these events are upgoing.

Also shown in Figure~\ref{fig:results} in a solid line is the result
of the Monte Carlo prediction, assuming no oscillations.  Integrating
over all zenith bins, the Monte Carlo predicts, for B-Up-Down, $159
\pm 40_{theor}$ events and for CT-Up, $144 \pm 36_{theor}$.  Forming
the ratio of observed to expected, we get our preliminary results,

\vskip 1 em
$R_{B-Up-Down} = 0.75 \pm 0.07_{stat} \pm 0.08_{sys} \pm 0.19_{theor}$

\vskip 1 em
$R_{CT-Up} = 0.59 \pm 0.06_{stat} \pm 0.06_{sys} \pm 0.15_{theor}$

\vskip 1 em
\noindent
Both ratios, especially CT-Up, differ significantly from unity, though
the significance is greatly degraded by the theoretical uncertainty,
of order 25\%, on the neutrino flux and neutrino cross section.  We
have not done a complete analysis of oscillations pending a more
careful calculation of the uncertainties.  However, the zenith
distributions are highly suggestive of oscillations with maximal
mixing and $\Delta m^2$ of a few times $10^{-3}$.  With these
parameters, we expect neutrinos from below, which travel thousands of
kilometers through the earth, to be fully oscillated (50\% deficit)
while neutrinos from above or from the horizontal, which travel only
tens of kilometers, are hardly suppressed at all.  For the B-Up-Down
analysis, in which the vertical bins sum over upward and downward
neutrinos, we expect half the upward and none of the downward
neutrinos to disappear, giving a deficit of 25\%.  In fact, we have
shown on Figure~\ref{fig:results} in the dashed lines the expectation
for a test point of $sin^2 2\theta = 1$ and $\Delta m^2 = 2.5 \times
10^{-3}$ which seems to fit the data quite well (but the remarkable
fit should not be taken at face value due to the large theoretical
errors in the prediction, not shown in the figure).

\section{Conclusions}

The CT-Up and B-Up-Down analyses, presented in preliminary form here,
provide a measurement of the atmospheric neutrino flux that is
independent of, and complementary to, the MACRO TG analysis, although
with smaller statistics.  Both preliminary analyses seem inconsistent
with no oscillations, although the theoretical uncertainties are
large.  Both analyses agree well with oscillations with parameters
($sin^2 2\theta = 1$ and $\Delta m^2 = 2.5 \times 10^{-3}$) consistent
with those suggested by the MACRO TG analysis, as well as other
experiments.

In the near future, we will analyze an additional 1.25 years of data
already in the can, and do further work to quantify systematic and
theoretical uncertainties.  We will also compute double ratios to try
to cancel some uncertainties.  A ratio of the present low-energy
events to the standard throughgoing events is of limited value in
canceling errors, because the primary cosmic ray flux and the cross
sections at the different energies may have different systematics.  A
more promising ratio is that of the two low-energy analyses,
$\frac{R_{B-Up-Down}}{R_{CT-Up}}$.  Referring again to
Figure~\ref{fig:scnrg}, these events come from the same energy range,
so a great cancellation of uncertainties will occur.  This should
strengthen the conclusions outlined in rough form in
Section~\ref{sec:results}.

\end{document}